\documentclass[12pt,preprint]{aastex}

\usepackage{epsfig}
\usepackage{amsmath}
\usepackage{euscript}

\shorttitle{Neutrino Heating Between Neutron Stars}
\shortauthors{Salmonson and Wilson}

\received{2002 March 20}
\begin{document}

\title{A Model of Short Gamma-Ray Bursts: Heated Neutron Stars in Close Binary Systems}

\author{Jay D. Salmonson} 
\author{James R. Wilson} 
\affil{Lawrence Livermore National Laboratory, P.O. Box 808, Livermore, CA 94550}

\begin{abstract}
In this paper we present a model for the short ($< 1$ second)
population of gamma-ray bursts.  In this model heated neutron stars in
a close binary system near its last stable orbit emit a large amount
of neutrinos ($\sim 10^{53}$ ergs).  A fraction of these neutrinos
will annihilate to form an $e^+e^-$ pair plasma wind which will, in
turn, expand and recombine to photons which make the gamma-ray burst.
We study neutrino annihilation and show that a substantial fraction
($\sim 1/2$) of energy deposited into $e^+e^-$ pairs comes from
inter-star neutrinos, where each member of the neutrino pair
originates from each neutron star.  Thus, in addition to the
annihilation of neutrinos blowing off of a single star, there is a new
source of baryon-free plasma that is deposited between the stars.  To
model the $e^+e^-$ pair plasma wind between stars, we do
three-dimensional relativistic numerical hydrodynamic calculations.
We find that the time scale for these bursts, deriving from the
baryon-free plasma, is less than one second and they will have a hot
spectrum $\sim 5$ MeV.  The energy in bursts is the order of $10^{52}$
ergs.
\end{abstract}


\section{Introduction}

In \citet{swm01} we investigated a model for gamma-ray bursts (GRBs)
deriving from a neutrino burst of energy $\sim 10^{53}$ ergs above a
heated, collapsing neutron star in a binary.  Such conditions have
been suggested by numerical relativistic hydrodynamic simulations
\citep{mw00} of compression, heating and collapse of binary neutron
stars near their last stable orbit.  Such a thermal neutrino burst was
found to partially recombine via $\nu\overline{\nu} \rightarrow
e^+e^-$ into an $e^+e^-$ pair plasma which expands relativistically.
This fireball then recombines into photons via $e^+e^- \rightarrow
\gamma\gamma$ and was found to give a gamma-ray burst of energy $\sim
10^{51} - 10^{52}$ ergs with spectral and temporal characteristics
consistent with observations.

This paper elaborates on the previous work in three distinct ways. 1)
The latest general relativistic hydrodynamic calculations by Wilson
and Mathews indicate that evolution and compression of the stars in a
binary is faster than previously thought; $\lesssim 1$ sec, thus
making this process a more natural candidate for the generation of
``short'' GRBs ($\lesssim 1$ sec). 2) Herein we use recent work by
\citet{sw01} which calculates the neutrino annihilation rate {\it
between} two neutron stars in addition to the annihilation rate from a
single star \citep{sw99}.  This `inter-star' annihilation will not
drive a baryon wind from the stellar surface and thus we find no
baryon loading in the $e^+e^-$ pair plasma that originates between the
neutron stars.  3) We employ three dimensional (3D) relativistic
hydrodynamic calculations to model the flow of the wind in the
complex, rotating, strong gravity environment around the neutron
stars.

It has been known for some time that the population of GRBs is
bimodal, with approximately a third of bursts having durations less
than 2 seconds, and these short bursts typically have harder spectra
than those bursts lasting longer than 2 seconds \citep{kmfb+93}.  This
bimodality is thought to indicate that short and long bursts are
separate populations and mechanisms.  This view is bolstered by
evidence that spectral break energy \citep{ppbm01} and pulse lags
\citep{nsb00} have a discontinous jump between short and long
populations.  The lack of discovery and location of an afterglow
associated with short bursts has severely impeded progress in
determining the distance scale and thus the energetics of these
bursts. In fact, comparisons of BeppoSAX and BATSE data archives
indicates a dearth of X-ray afterglows \citep{g+01}.  Analysis of
BATSE data for short burst decay tails give conflicting results
indicating both a lack of an underlying afterglow component
\citep{c00} and the existence of such a component \citep{lrrg01}.

\section{The Model}

In this model we estimate that 10\% of the rotationally enhanced
binding energy of a neutron star ($\sim 10^{53}$ ergs) is converted to
thermal energy via compression, vortices and shocks.  This $10^{53}$
ergs of energy is released as a monotonically increasing luminosity of
neutrinos over a timescale of order $\sim 1/10$ second as estimated by
\citet{mw97}.  This neutrino luminosity $\sim 10^{54}$ ergs s$^{-1}$
annihilates into an $e^+e^-$ pair plasma.  Annihilation of high
neutrino luminosities in the strong gravitational field of the neutron
stars can have high efficiencies; near unity \citep{swm01}.  About
half of the pair plasma energy is deposited uniformly around the
neutron stars due to single star neutrino annihilation \citep{sw99}
and the other half is deposited between the stars due to interstar
neutrino annihilation \citep{sw01}.

\begin{figure}
\plotone{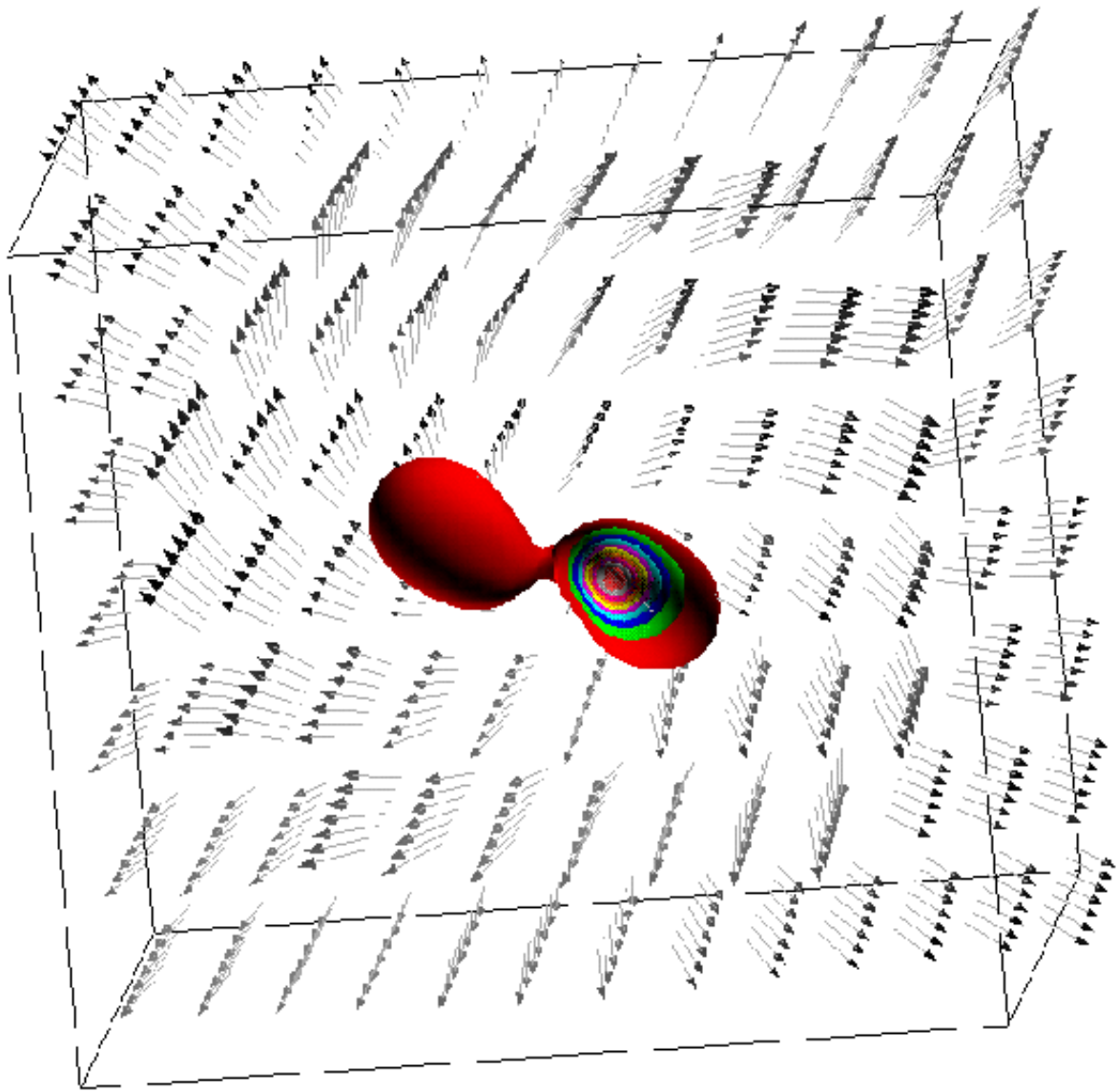}
\caption{Three dimensional relativistic simulation of two 10 km radius
neutron star separated by 30 km emitting $10^{53}$ ergs/sec of energy
in $e^+e^-$ pair plasma and about 1 \% equivalent mass in baryons.
The contour map, with right star cutaway, is of baryon density. The
vector field is the 3-velocity of expanding plasma.  This problem
settles down to a static flow after about one orbit with period $\sim
1/300$ second.  \label{fig:3Dfig} }
\end{figure}

This plasma deposition morphology then becomes input for the 3D
relativistic hydrodynamic code, which calculates the expansion of the
plasma (Figure \ref{fig:3Dfig}).  These simulations show a plasma of
very high entropy expanding out along the plane of symmetry between
the neutron stars.  In the regions around the stars lower entropy
plasma is formed because a baryon wind is blown from the stars.  The
numerical model follows the baryons, hence some baryons were added in the
interstellar region.  The calculation yields an estimate of the
angular distribution of the baryonic loading of the plasma.

Thus this model predicts a variety of bursts.  Viewed along the axis
of rotation, a prompt quasi-thermal burst of duration $\sim 1/10$
second will result from the annihilation of fireball pairs
\citep{swm01}.  Because of the dearth of baryons left over to sweep
into the interstellar medium, we do not predict the existence of an
afterglow.  This agrees with preliminary searches of the data archives
for short-burst afterglows, which appear to be missing \citep{g+01}.

Viewed far from the axis of rotation a very different burst results.
The lower entropy means that there will not be a prompt burst from
pair annihilation in the fireball.  However, there will be a baryon
wind sweeping into the interstellar medium.  Thus we expect a burst
that decays into an afterglow as a power-law \citep{swm01}.  This
behavior will be made chaotic and complex by the rapid rotation of the
binary system.

\section{One-Dimensional Calculations \label{1dsection}}

Using one-dimensional (1D) relativistic hydrodynamics described in
\citet{swm01}, we can study the expansion of an $e^+e^-$ pair plasma
fireball from the surface of a single neutron star.  As per our model,
we deposit $\EuScript{E} = 10^{52}$ ergs of energy into a narrow
(width $= 1$ km) shell region above the surface of a $R_0 = 10$ km
radius and 1.4 $M_\odot$ mass neutron star.  We monotonically ramp the
energy over $\Delta t = 0.1$ second.  This model is run with a range
of deposited baryon densities and thus a range of baryon loadings of
the fireball.  

It is important to note that our hydrodynamics do not track the
protons and neutrons separately and, as such, assumes they are well
coupled during the expansion and acceleration of the plasma.  As was
shown by \citet{fpa00} and \citet{dkk99}, the neutrons can decouple
from the protons and $e^+e^-$ plasma during acceleration, thus
changing the overall baryon load of the plasma.  Since it is plausible
that the baryon matter blown off of the neutron star is 90\% neutrons,
it is reasonable to expect, depending on the efficiency of the
decoupling, the effective baryon mass to be a factor of several less
than the total baryon mass.

\begin{figure}
\epsscale{.8}
\plotone{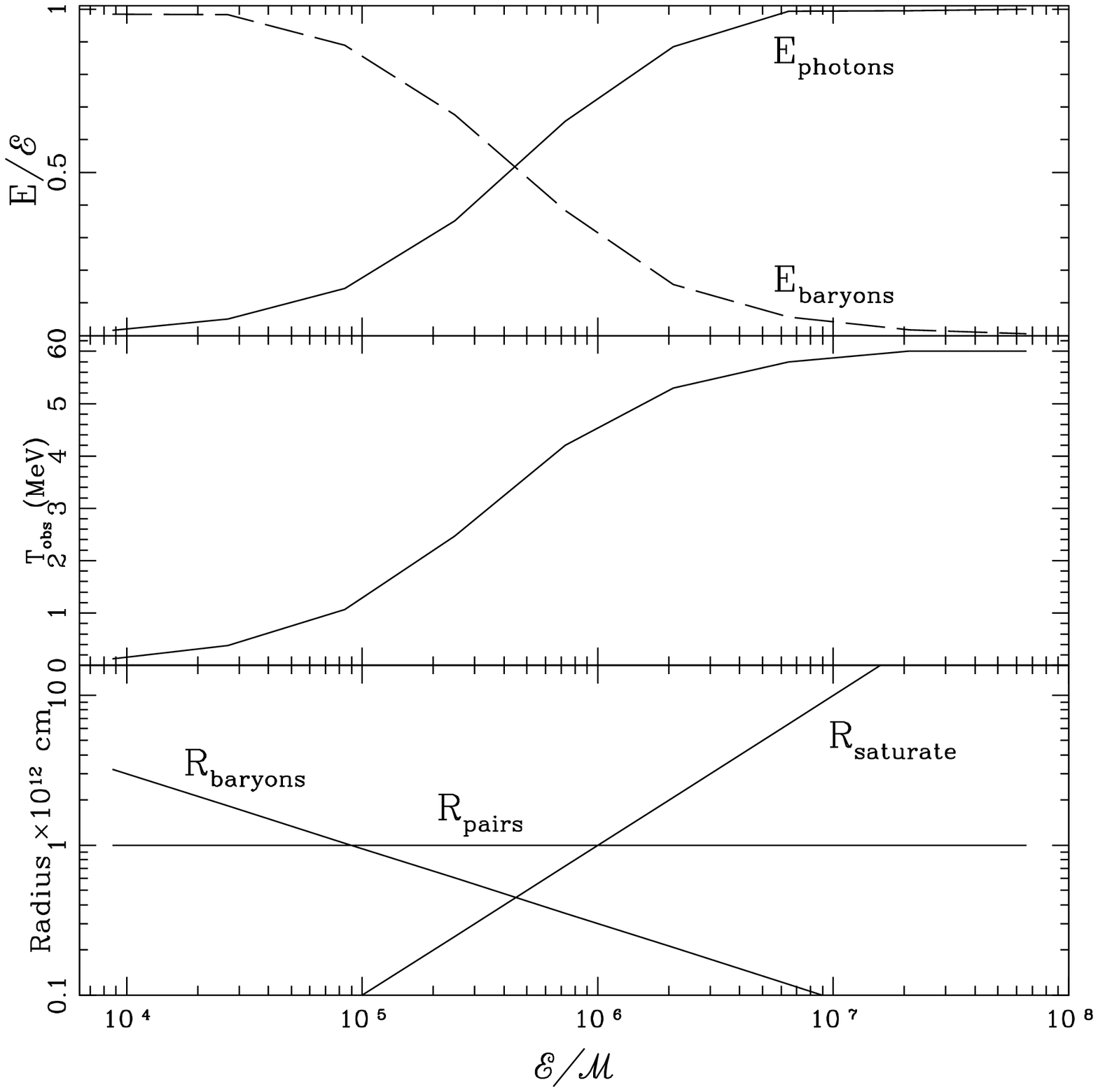}
\caption{Output of the 1D code run over a range of initial
energy-to-mass ratios, $\EuScript{E/M}$, with energy fixed at
$\EuScript{E} = 10^{52}$ ergs. This shows that the energy available
for creation of an internal burst versus that for an external burst
varies as a function of baryon loading.  The more baryon loading
(lower $\EuScript{E/M}$), the more energy is entrained in the baryons,
$E_{\text{baryons}}$, and thus creates a more energetic external
shock.  Less baryon loading (higher $\EuScript{E/M}$) makes a more
efficient fireball burst, where much of the original energy is left in
the photons, $E_{\text{photons}}$. The middle figure shows the typical
observed temperature, $T_{\text{obs}}$, of the resultant photons.  The
lower figure shows the characteristic radii determining the evolution:
$R_{\text{baryons}} \approx 10^{12} \sqrt{(\EuScript{M/E})_{-5}}$ cm,
$R_{\text{pairs}} \approx 10^{12}$ cm, $R_{\text{saturate}} \approx
10^6 (\EuScript{E/M})$ cm.  The photons decouple from the baryons at
the maximum, $R_{\text{baryons}}$ or $R_{\text{pairs}}$.  If
$R_{\text{saturate}}$ is greater than either $R_{\text{baryons}}$ or
$R_{\text{pairs}}$, then the baryons are unable to saturate with
energy and thus the energy remains in the photons.
\label{fig:eod}}
\end{figure}

The plasma fireball is evolved and we find that it expands
adiabatically as a thin shell of constant coordinate thickness $\Delta
R$ \citep{rswx00}.  We track this expansion and pair annihilation
until the optical depth, $\tau$, of the plasma falls below unity:
$\tau \sim N \sigma_T \Delta R < 1$ where $N$ is the coordinate
density of electrons and positrons and $\sigma_T$ is the Thompson
cross-section.  The energy in photons, $E_{\text{photons}}$, and in
the baryons, $E_{\text{baryons}}$, is calculated and these results are
summarized in Fig.~(\ref{fig:eod}).  One can see that for
$\EuScript{E/M} \lesssim 10^5$, where $\EuScript{M}$ is the baryonic
rest mass energy, most of the energy goes into accelerating the
baryons while for $\EuScript{E/M} \gtrsim 10^5$ a substantial fraction
of the energy is expressed as photons directly from the plasma; i.e.~a
gamma-ray burst.

This result can be understood when one considers that the opacity of
the fireball derives from two electron densities: electrons associated
with entrained protons, $N_{electrons} = N_{protons}$, and electrons
and positrons in pairs, $N_{pairs}$.  Defining the optical depth as
$\tau = \tau_{electrons} + \tau_{pairs} \approx (N_{protons} +
N_{pairs}) \sigma_T \Delta R$, then by comparing the densities
$N_{protons}$ and $N_{pairs}$, we understand the relative importance
of these two sources of opacity.

We first consider the baryons.  Equating the energy deposition with
outgoing flux gives the initial energy density at the surface of the
neutron star $E_0 = \EuScript{E}/(\Delta t 4\pi R_0^2 c) \approx 3
\times 10^{29}$ ergs cm$^{-3}$.  It follows that the corresponding
initial number density of baryons is $N_{b,0} = E_0/m_p c^2
(\EuScript{M/E}) \approx 1.3 \times 10^{35} (\EuScript{M/E})
\text{cm}^{-3}$, where $\EuScript{M/E}$ is the ratio of baryon rest
mass energy to total energy.  Finally, we know \citep{rswx00} that for
relativistic adiabatic expansion we have: $N_b = N_{b,0} (R_0/R)^2$.
Let us assume that 10\% of the baryons are protons with associated
electrons.  Then the optical depth from the entrained baryons is
$\tau_{electrons} \approx 0.1 N_b \sigma_T \Delta R \sim 0.1 (E_0/m_p
c^2)(\EuScript{M/E})(R_0/R)^2 \sigma_T \Delta R$ and thus optical
thinness, $\tau_{electrons} \approx 1$ happens at a radius
$R_{baryons} \sim 10^{12} \sqrt{(\EuScript{M/E})_{-5}}$ cm, where
$(\EuScript{M/E})_{-5} \equiv (\EuScript{M/E})/10^{-5}$.  Notice that
this radius depends on the initial baryon loading of the fireball:
$R_{baryons} \propto \sqrt{\EuScript{M/E}}$.

The opacity due to pairs is a bit more complex.  As discussed in
\citet{swm01}, the pair number density is governed by the following
equation
\begin{equation}
\frac{\partial N_{pairs} }{ \partial t} = - \frac{\alpha }{ r^2} \frac{
\partial }{ \partial r} (\frac{r^2 }{ \alpha} N_{pairs} V^r) +
\overline{\sigma_{ann} v} ((N_{pairs}^0 (T'))^2 - N_{pairs}^2)/W^2 
\label{jay:E:ndiff}
\end{equation}
where $\overline{\sigma_{ann} v}$ is the Maxwellian averaged mean pair
annihilation rate per particle and $N_{pairs} (T')$ is the local
coordinate equilibrium $e^+e^-$ pair density at local temperature $T'$
given by the appropriate Fermi integral with a chemical potential of
zero.  Early in the fireball expansion $e^+e^-$ pairs are in
equilibrium so $N_{pairs} \approx N_{pairs}^0(T')$ and thus the pairs
expand adiabatically: $N_{pairs} = N_{pairs,0} (R_0/R)^2$, where the
initial pair number density is given by $N_{pair,0} \sim (a
E_0^3)^{1/4}/3k \sim 10^{34}~\text{cm}^{-3}$.  When the local
temperature, $T'$, of the plasma falls below $T'_\ast \simeq m_e c^2/3
\sim 1/6$ MeV the pairs begin to rapidly annihilate.  This happens at
a radius $R_\ast \sim R_0/T' (E_0/a)^{1/4} \sim 100 R_0$.  The pair
number density plummets until the annihilation term in
Eqn.~(\ref{jay:E:ndiff}) no longer dominates, which happens at a
density $N_{pair,\ast} \sim R_\ast/(\overline{\sigma_{ann}} R_0^2)
\sim 10^{20} \text{cm}^{-3}$.  This annihilation happens over a few
$R_\ast$ which is very small compared to the radius at which pairs and
photons decouple and therefore these results are independent of the
details of this annihilation phase.  Beyond this point the pair number
density again expands adiabatically: $N_{pair} = N_{pair,\ast}
(R_\ast/R)^2$.  Thus the optical depth from the pairs is $\tau_{pairs}
\approx N_{pairs} \sigma_T \Delta R \sim (E_0/a)^{1/4} \sigma_T \Delta
R/(\overline{\sigma_{ann}} R_0 {T'}^3_\ast) (R_0/R)^2$ and optical
thinness, $\tau_{pairs} \approx 1$, happens at a radius $R_{pairs}
\sim \sqrt{(E_0/a/T'_\ast)^{3/4} \sigma_T R_0 \Delta
R/\overline{\sigma_{ann}}} \sim 10^{12}$ cm which is independent of
$\EuScript{E/M}$.

Thus we see that the transition shown in Fig.~(\ref{fig:eod}) from
baryon-dominated to photon-dominated fireball energy is governed by
the relative importance of opacity associated with the baryons and
that associated with the $e^+e^-$ pairs.  Therefore it is important to
consider both the opacity due to electrons associated with protons,
which dominate for high baryon loading, and the opacity of the
$e^+e^-$ pair plasma which dominates for low baryon loads.  For
initial values $\EuScript{E/M} < 10^5$, $R_{\text{baryons}} >
R_{\text{pairs}}$ so the opacity is dominated by the electrons
associated with the baryons and thus the baryons are able to ``hold''
the plasma longer and thus absorb all of the fireball energy.  For
$\EuScript{E/M} > 10^5$ we have $R_{\text{pairs}} >
R_{\text{baryons}}$ and the pairs are the dominant source of opacity
and thus determine the radius of optical thinness.  If the radius of
optical thinness falls below the radius to saturate the energy of the
baryons, i.e.~the radius at which most of the plasma energy is in the
baryons, $R_{\text{saturate}} = (\EuScript{E/M}) R_0$, then the
acceleration of the baryons is inefficient, they do not attain the
energy $(\EuScript{E/M}) m_p c^2$, and thus more energy remains in the
photons, thus yielding a photon burst.

\section{Three-Dimensional Calculations }

While the 1D hydrodynamics of the last section elucidate the behavior
of an expanding relativistic plasma, a 3D calculation is required to
capture the structure inherent in the present problem of rotating,
emitting binary neutron stars.  In this section we introduce the 3D
relativistic hydrodynamic equations.  Herein we use units $G = c =
1$. The continuity equation is
\begin{equation}
\frac{\partial D }{ \partial t} = -6 D \frac{ \partial \log \phi }{ \partial t} - \frac{1}{\phi^6} \frac{ \partial ({\phi^6 D V^i}) }{ \partial x^i} ~.
\label{eqn:Ddens}
\end{equation}
where the spatial dimensions are indicated by the sum over $i =
1,2,3$.  The energy equation is
\begin{equation}
\frac{\partial E }{ \partial t} = - 6 \Gamma E \frac{ \partial \log \phi}{\partial t} - \frac{1}{\phi^6} \frac{ \partial ({\phi^6 E V^i}) }{ \partial x^i }  - P \biggl[ \frac{\partial W }{ \partial t} +
\frac{1}{ \phi^6} \frac{\partial ({ W V^i \phi^6}) }{ \partial x^i} \biggr] ~.
\label{eqn:Edens}
\end{equation}
And the momentum equations are
\begin{equation}
\begin{split}
\frac{\partial S_i }{ \partial t} = &- 6 S_i \frac{ \partial \log \phi}{\partial t} - \frac{1}{\phi^6} \frac{ \partial {(\phi^6 S_i V^j)} }{
\partial x^j}  - \alpha \frac{\partial P }{ \partial x^i} - W (D + \Gamma E) \frac{ \partial \alpha }{ \partial x^i } \\ 
&+ S_j \frac{ \partial \beta^j }{ \partial x^i } + 2 \alpha (D + \Gamma E) \biggl[W - \frac{1}{W} \biggr] \frac{ \partial ( \log \phi) }{\partial x^i} ~,
\end{split}
\end{equation}
where the four-velocity is 
\begin{equation}
U_i = \frac{S_i }{ D + \Gamma E} ~,
\label{fourveleqn}
\end{equation} 
and we define a generalized Lorentz factor
\begin{equation}
W = \sqrt{ 1 + \frac{\Sigma (U_i)^2}{\phi^4}} ~,
\label{eqn:W}
\end{equation}
the coordinate velocity in terms of the four velocity
(eqn.~\ref{fourveleqn}) is given by
\begin{equation}
V^i = \frac{\alpha U_i }{W \phi^4} - \beta^i ~,
\end{equation}
and the equation of state is given by 
\begin{equation}
\Gamma = 1 + \frac{ P W }{ E} ~.
\end{equation}

The  metric  is 3+1  and  taken  to  be  conformally flat.  The  lapse
function is given by
\begin{equation}
\alpha \equiv \Biggl( \frac{1 - \frac{F}{2}}{1 + \frac{F}{2}} \Biggr) ~.
\label{eqn:alpha}
\end{equation}
The spatial factor is
\begin{equation}
\phi \equiv 1 + \frac{F}{2} ~,
\end{equation}
and the shift vector is
\begin{equation}
{ \beta^\phi} \propto {\bf \Omega} \times {\bf r} - A \Biggl(\frac{M_1}{r_1} - \frac{M_2}{r_2}\Biggr) ~.
\end{equation}
The factor $F$ is defined by
\begin{equation}
F \equiv 
\begin{cases}
\frac{M}{r_1} + \frac{M}{r_2}&  r_1 > R_\star \\
\frac{M}{2} \Biggl(\frac{3}{R_\star} - \frac{r_1^2}{ R_\star^3} \Biggr) + \frac{M}{r_2}& r_1 < R_\star
\end{cases}
\end{equation}
where $r_1$ and $r_2$ are the distances from the center of the nearest
and farthest stars respectively, each of mass $M$ and radius
$R_\star$.

\begin{figure}
\epsscale{1.0}
\plotone{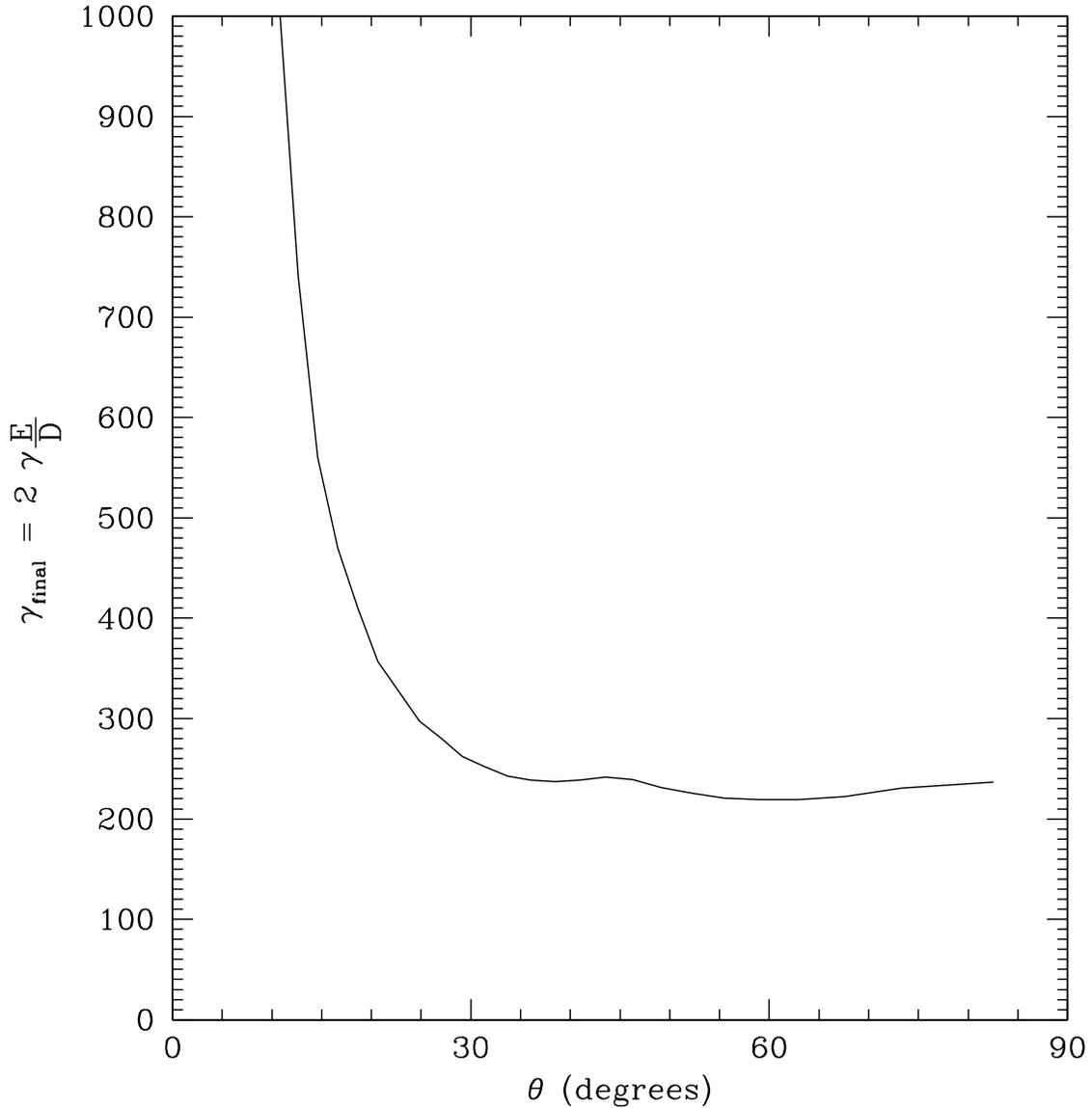}
\caption{The terminal (final) Lorentz factor, $\gamma_{\text{final}} =
2 \gamma E/D$, for $\gamma,E/D \gg 1$, from 3D simulations as a
function of angle from the rotation axis of the binary system in the
plane described by said axis and the line connecting the two stars.
The local Lorentz factor is $\gamma$.  Thus $0^\circ$ is along the
rotation axis and $90^\circ$ corresponds to being directly behind one
of the stars. The most energetic material, $\gamma_{\text{final}}
\rightarrow \infty$, is located within a narrow region, $\theta <
15^\circ$, on the symmetry plane between the two stars.
\label{fig:EoD} }
\end{figure}


\begin{figure}
\plotone{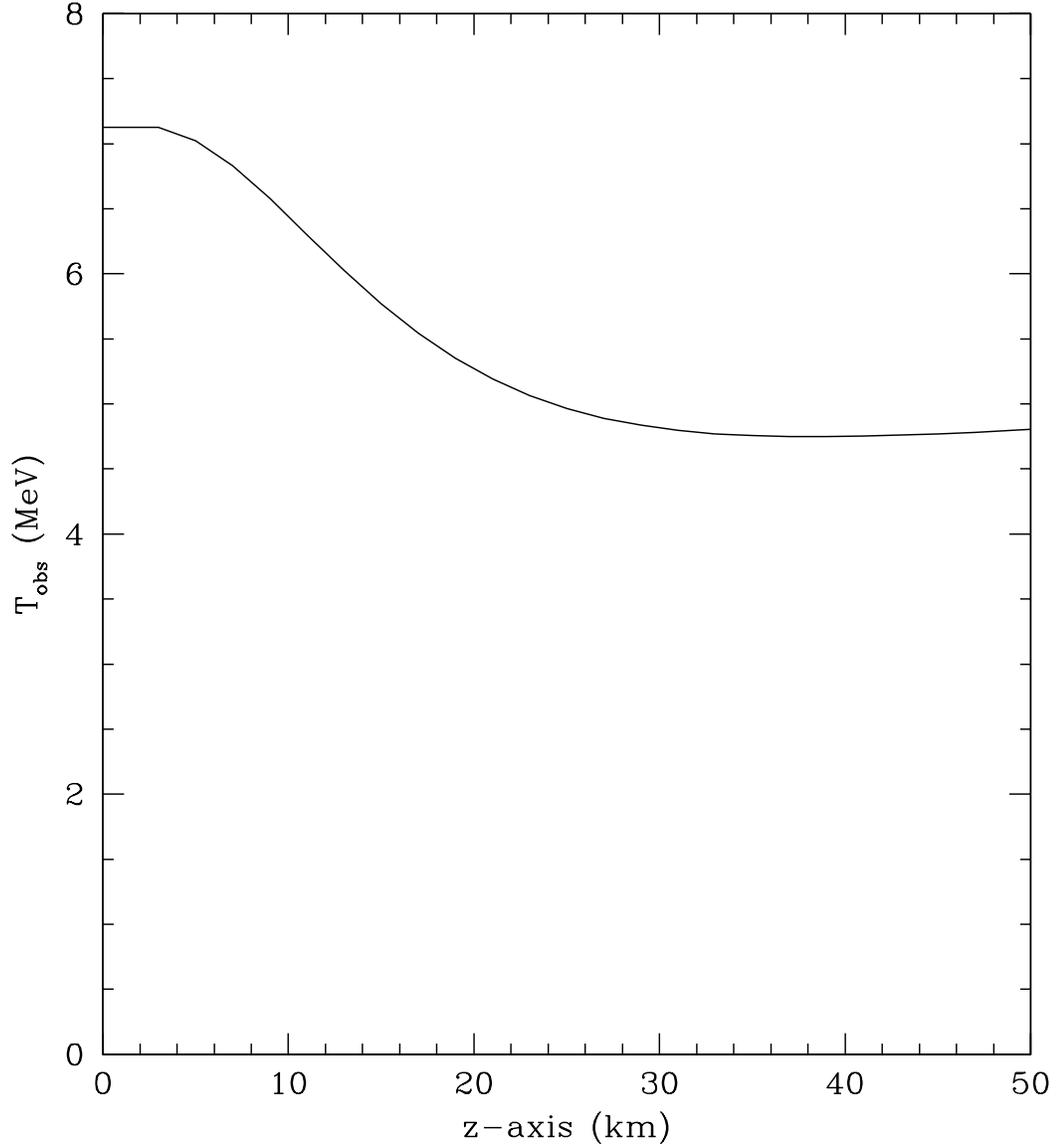}
\caption{From the 3D code, the observed temperature at infinity,
$T_{obs} \propto \alpha ( W^3 E)^{1/4}$ of the baryon-free pair plasma
along the z-axis of rotation. Most of the plasma is at the high value
of z hence this observed average effective temperature will be near 5
MeV.
\label{fig:Tobs} }
\end{figure}

\section{ Results \label{sec:gammaschem}}

An example of a 3D run is shown in Fig.~\ref{fig:3Dfig}.  An edit of
the final Lorentz factor, $\gamma_{\text{final}}$, as a function of
angle from the rotation axis in the plane defined by the rotation axis
and the line connecting the two stars is shown in Fig.~\ref{fig:EoD}.
One can see that the highest energy per baryon occur along the
rotation axis.  From this figure one gets the scale of the opening
angle of this high energy material: $\sim 15^\circ$.  The simulation
does not sufficiently resolve the region of high $E/D$ at low angles.
Very little baryon mass will make it into this region of high energy
density, thus we expect $E/D$ to be effectively infinite on the
rotation axis.  Thus for viewers within this angle we have a thermal
burst dominated by $e^+e^-$ pair-photon plasma as described in
\citep{swm01}.  An edit of the observed temperature $T_{obs} \propto
\alpha (W^3 E)^{1/4}$ is shown in Fig.~\ref{fig:Tobs}, where we have
used quantities defined in eqns.~(\ref{eqn:Edens}, \ref{eqn:W},
\ref{eqn:alpha}).

One can see from Fig.~\ref{fig:EoD} that, depending on the observer
angle with respect to the axis of rotation, a wide range of baryon
loadings, parameterized by $E/D$, will be observed.  Thus we study the
range of expected observed signals as it depends on $E/D$.  Herein we
assume a correspondence between the ratio of energy to mass,
$\EuScript{E/M}$, of the spherically symmetric calculations of Section
\ref{1dsection} to the ratio of energy to mass {\it coordinate
densities}, $E/D$, as taken from the 3D code runs.  As such, we expect
a thermal burst for observers along the axis of rotation of the stars
and within an angle $\sim 15^\circ$ where no baryons are deposited and
thus $E/D \rightarrow \infty$.  As seen in Fig.~\ref{fig:EoD}, for
increasing $\theta$ from the rotation axis, $E/D$ decreases rapidly to
$\sim 100$.

For all $E/D = \EuScript{E/M} < 10^5$ all of the energy of the pair
plasma is deposited into the baryons and therefore a prompt thermal
burst does not occur.  Instead, the mass, $\EuScript{M}$, of baryons
moving with Lorentz factor, $\gamma_{\text{final}} = \EuScript{E/M}$, sweep
into the interstellar medium, thus forming a collisionless shock, and
decelerate according to \citep[e.g.][]{piran00}
\begin{equation}
\EuScript{E} = \gamma_{\text{final}} \EuScript{M} = 4 \pi \rho \gamma^2 r^3
\end{equation}
for
\begin{equation}
r > r_{dec} \equiv \biggl(\frac{\EuScript{E}}{4 \pi \rho c^2 \gamma_{\text{final}}^2}
\biggr)^{1/3} \sim 10^{16} \gamma_{300}^{-2/3} E_{52}^{1/3} n_1^{-1/3} \quad
\text{cm}
\end{equation}
where $r_{dec}$ is the deceleration radius, $\rho$ is the interstellar
medium (ISM) density, $n_1 \equiv \rho/m_p$ cm$^{-3}$ is the ISM
baryon number density, $\gamma_{300} \equiv \gamma/300$ and $E_{52}
\equiv \EuScript{E}/(10^{52}~\text{ergs})$.  This model was worked out
in detail in \citet{swm01}.  The shock will emit synchrotron radiation
as it decelerates with a characteristic luminosity $L \sim (n_1
\gamma_{300}^8)^{1/3} E_{52}^{2/3} 10^{51}$ ergs s$^{-1}$ at a
characteristic photon energy $\epsilon \sim 200 \sqrt{n_1}
\gamma_{300}^4$ keV and a duration on the deceleration timescale
$t_{dec} = r_{dec}/2\gamma^2 c \sim 3~ (E_{52}/n_1/
\gamma^8_{300})^{1/3}$ s.  Note that for $\gamma \lesssim 300$ this
timescale is consistent with long bursts: $> 3$ s.

\begin{figure}
\plotone{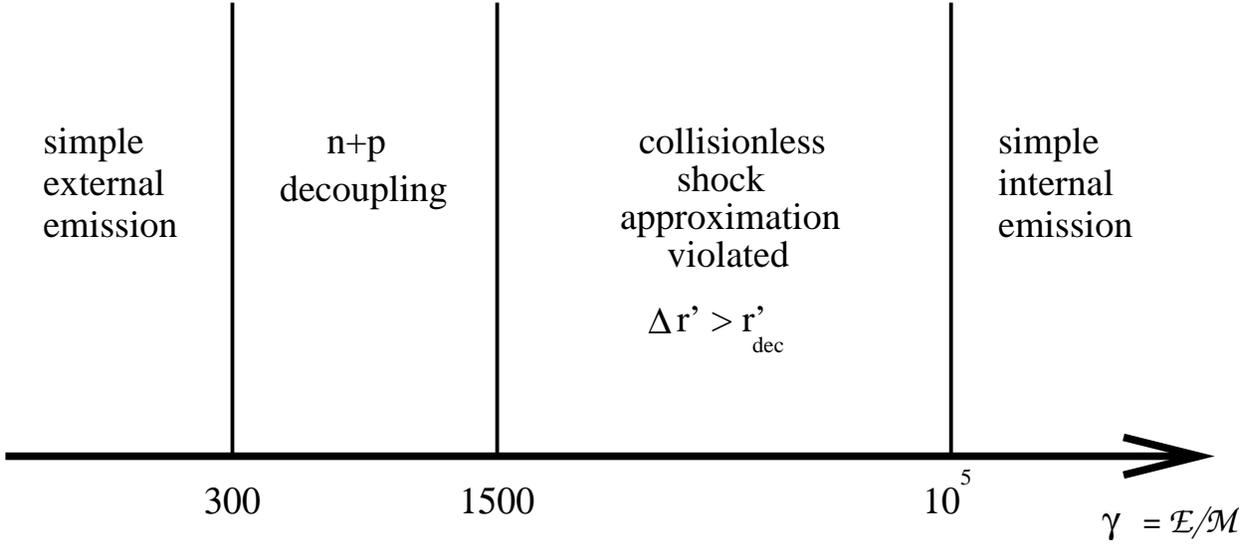}
\caption{A schematic figure representing the physical regions along
the spectrum of $\gamma = \EuScript{E/M}$ discussed in Section
\ref{sec:gammaschem}.  A simple external shock results for $\gamma <
300$.  For $300 < \gamma < 1500$ the neutrons will be decoupled from
the protons and thus will decay and shock at a larger radius, thus
creating a smeared external shock.  For $1500 < \gamma < 10^5$ the
external shock cannot be considered a simple thin shock and the theory
of collisionless shocks is likely violated. In the region $\gamma >
10^5$, the energy remains in the photons from the original pair plasma
and thus a simple internal shock results as discussed in Section
\ref{1dsection} and is detailed in Fig.~(\ref{fig:eod}).
\label{fig:gammaschem}}
\end{figure}

A distinct feature of this progenitor model, since the wind comes from
the surface of a neutron star, is that most of the baryons will be in
the form of neutrons (say 90\%).  Thus these neutrons will not couple
with the ISM magnetic field and will pass by the decelerating protons.
These neutrons will decay back to protons on a timescale of a neutron
half-life, $\tau_{n,1/2} \approx 1000$ s, which happens at a radius
$R_{decay} = \gamma c \tau_{n,1/2} \sim \gamma_{300} 10^{16}$ cm.
Comparing this radius with the decelaration radius, $r_{dec}$, we see
that $r_{dec} > R_{decay}$ for $\gamma < 300$, i.e.~the neutrons have
decayed into protons before reaching the deceleration radius.  So we
expect this effect to be important for $\gamma = \EuScript{E/M} >
300$ (Fig.~\ref{fig:gammaschem}).

Another characteristic Lorentz factor $\gamma = \EuScript{E/M}$ is
when the time, in the fluid frame, to reach the deceleration radius is
shorter then the deposition time, $\Delta t = 0.1$ s.  The fluid frame
deposition distance is $\Delta r' = \gamma c \Delta t \sim
\gamma_{300} 10^{12}$ cm.  Also in the fluid frame, the deceleration
distance is $r'_{dec} = r_{dec}/\gamma \sim 10^{15} (n_1
\gamma_{300}^5)^{-1/3}$ cm.  These are equal at $\gamma \approx 1500$
(Fig.~\ref{fig:gammaschem}).  Thus for $\gamma > 1500$ the plasma
deposited at the beginning of the deposition time has begun to
decelerate before the plasma has completed deposition, so the
assumption that the plasma is a thin, shocked shell plowing into the
ISM is invalid.  Instead, energy will be continuously deposited into
the shock.

An important caveat for these high Lorentz factors, $\gamma \gg 300$,
is that they likely are beyond the regime of validity for the
collisionless shock.  The detailed physics behind such shocks is not
well understood \citep[e.g.][]{piran00} and assumes that the kinetic
energy of the baryons, $(\gamma-1) m_p c^2$, will effectively amplify
the magnetic field and energize the electrons.  However, low densities
of ultra-high energy baryons will inefficently couple with the ISM and
thus the collisionless shock approximation will be invalid.  

Again, for energy, $\EuScript{E} = 10^{52}$ ergs, with photon energy
$\epsilon \approx 200 \gamma_{300}^4$ keV, the observed photon fluence
seen at a distance, $R$, will be $\EuScript{E} \xi /4 \pi R^2 \epsilon
\approx 250 \xi \gamma_{300}^{-4} R_{Gpc}^{-2}$ cm$^{-2}$, where
$R_{Gpc} \equiv R/(1~ \text{Gigaparsec})$ and $\xi$ is an efficiency
factor which may be about $10\%$.  For example, given $\gamma =
1500$, we expect a fluence of $0.04$ photons cm$^{-2}$ with
characteristic energy $\epsilon \approx 0.1$ GeV over a duration
$t_{dec} \approx 0.04$ s. So such a burst would be far from
recognizable as such (Fig.~\ref{fig:gammaschem}).

\section{Discussion}

A key result of the 3D numerical simulations (Fig.~\ref{fig:EoD}) is
that the emission from this system is bimodal: about half of the
total energy is deposited as a pure, baryon-free, $E/D \rightarrow
\infty$, pair plasma along a `fan' of angular half-width $\theta =
15^\circ$ along the symmetry plane between the neutron stars, and the
other half of the total energy is blown off of the neutron stars as a
wind with $E/D \approx 300$.  Very little of the energy is deposited
in intermediate regimes of $E/D$.  As such we analyse the expected
observations from this model.

\begin{figure}
\plotone{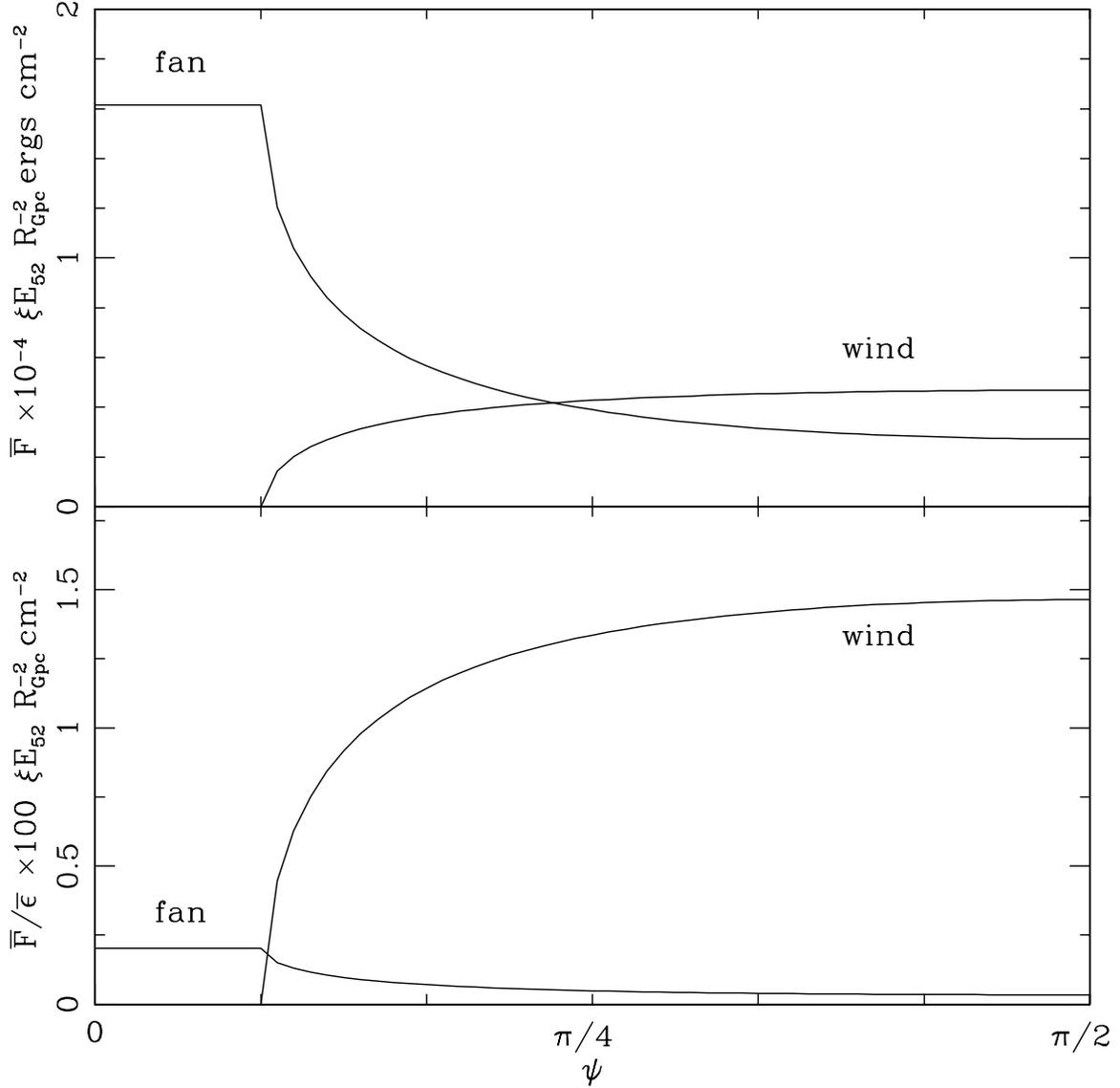}
\caption{ The top figure shows the average fluence $\overline{F}$ of
the fan (Eqn.~\ref{eqn:Ffan}) and the wind (Eqn.~\ref{eqn:Fwind}) as a
function of viewing angle $\psi$.  The bottom figure shows number
fluence, $\overline{F}$ divided by the characteristic photon energy
for the fan, $\overline{\epsilon}_{fan} \approx 5$ MeV
(Fig.~\ref{fig:Tobs}), and the wind, $\overline{\epsilon}_{wind}
\approx 200$ keV (Sec.~\ref{sec:gammaschem}).  This demonstrates that
most of the energy comes from the fan while most of the photons come
from the wind.  A viewer within $\psi < 15^\circ = \pi/12$ will see only the
fan emission, which effectively constitutes a jet of half-opening
angle $15^\circ$.}
\label{fig:fanwind}
\end{figure}

The solid angle subtended by the fan of baryon-free plasma is
$\Omega_{fan} = 4\pi\sin \theta$ and the wind blows off into the rest
of space, $\Omega_{wind} = 4\pi(1-\sin\theta)$.  Therefore the fluence
from each component at a distance $R$ for total energy $\EuScript{E}$ is
\begin{equation}
\begin{split}
F_{fan} &= \frac{\xi \EuScript{E}/2}{\Omega_{fan} R^2} = \frac{\xi \EuScript{E}}{8\pi\sin\theta R^2} \\
F_{wind} &= \frac{\xi \EuScript{E}/2}{\Omega_{wind} R^2} = \frac{\xi \EuScript{E}}{8\pi(1-\sin\theta) R^2} 
\end{split}
\end{equation}
where $\xi$ is an efficiency factor.  Now the system is rotating so an
observer with angle, $\psi$, from the rotation axis will see a mixture
of the fan and wind, unless he is located within an angle $\psi <
\theta$.  Since the period of rotation of the system is much smaller
than the energy deposition timescale, $\Delta t = 0.1$ s, we average
over the fluences from the wind and the fan to get a total average
fluence.  The proportion of a rotation subtended by the fan is $2/\pi
\arcsin(\theta/\sin\psi)$ for $\theta \lesssim \psi \leqslant \pi/2$
and is unity for $0 < \psi < \theta$.  So the average fluence
contributed by the fan as a function of viewing angle is (Fig.~\ref{fig:fanwind})
\begin{equation}
\overline{F}_{fan}(\psi) = \frac{\EuScript{E}}{8\pi R^2}
\begin{cases}
\frac{2/\pi \arcsin(\theta/\sin(\psi))}{\sin\theta}& \text{for
$\theta \lesssim \psi \leqslant \pi/2$} ~,\\
\frac{1}{\sin\theta}&  \text{for $0 < \psi < \theta$} ~,
\end{cases}
\label{eqn:Ffan}
\end{equation}
and that of the wind is
\begin{equation}
\overline{F}_{wind}(\psi) = \frac{\EuScript{E}}{8\pi R^2} 
\begin{cases}
\frac{1 - 2/\pi \arcsin(\theta/\sin(\psi))}{1 - \sin\theta}& \text{for
$\theta \lesssim \psi \leqslant \pi/2$} ~,\\ 0& \text{for $0 < \psi <
\theta$} ~.
\end{cases}
\label{eqn:Fwind}
\end{equation}
The total average fluence is
\begin{equation}
\overline{F}(\psi) = \overline{F}_{fan}(\psi) + \overline{F}_{wind}(\psi) ~.
\end{equation}
An observer within $0 < \psi < \theta$ sees only the fan emission.
This is effectively a jet of opening half-angle $\theta = 15^\circ$.
Such an observer sees a jet of strong thermal emission over a
timescale of 0.1 s.  For $\theta = 15^\circ = \pi/12$ we have
\begin{equation}
\begin{split}
\overline{F}(\psi = 0) &= 1.6 \times 10^{-4}~ \frac{\xi E_{52}}{R_{Gpc}^2} \quad \text{ergs cm}^{-2} \\
\overline{F}(\psi = \pi/2) &= 7.3 \times 10^{-5}~ \frac{\xi E_{52}}{R_{Gpc}^2} \quad \text{ergs cm}^{-2}
\end{split}
\end{equation}
so an observer along the jet ($\psi < \theta$) will see roughly twice
the fluence in a short, $0.1$ s, thermal, $T_{obs} \approx 5$ MeV,
gamma-ray burst, than an off-axis observer ($\psi \sim \pi/2$) who
will see a synchrotron burst of duration $\sim 3$ s with
characteristic photon energy 200 keV.

In conclusion, when the binary system is viewed within $15^\circ$ of
the axis of rotation, we have a model for short gamma-ray bursts,
$\sim 0.1$ s, which have hard spectra, several MeV.  In addition, when
the binary system is viewed at angles larger than $15^\circ$ from the
rotation axis, our model yields a second group of bursts, 30 times
more common, with soft spectra, a few hundred keV, and intermediate
time duration of 1 to 5 s.  The energy available depends on the
neutron star equation of state and the masses of the neutron stars.
The energy in a burst could range from $10^{50}$ ergs to a few times
$10^{52}$ ergs.  From Fig.~\ref{fig:fanwind} we see that if the
detector is primarily sensitive to number counts then at intermediate
viewing angles the fan contribution to the signal may be missed.  The
pure fan signal, i.e.~a jet viewed within $15^\circ$ of the rotation
axis, will have no afterglow, but the fan plus wind would have an
afterglow and a small contribution of high energy
photons. \citet{jfgh+01} have observed a 2 s burst with an afterglow.
Unfortunately the high energy detector data was not obtained.

This work was performed under the auspices of the U.S. Department of
Energy by University of California Lawrence Livermore National
Laboratory under contract W-7405-ENG-48.\\


\end{document}